\documentclass[
    ,final            
  ]
  {aipproc}
\layoutstyle{6x9}
\usepackage{graphicx,amsmath,wrapfig}
\begin{document}
\title{ Global analysis of AAC for determining \\
        polarized parton distribution functions}


\author{M. Hirai}{
  address={Institute of Particle and Nuclear Studies, 
          High Energy Accelerator Research Organization (KEK)},
   email={mhirai@post.kek.jp}}
   
\author{S. Kumano}{
  address={Institute of Particle and Nuclear Studies, 
          High Energy Accelerator Research Organization (KEK)},
  altaddress={Department of Particle and Nuclear Studies,
           The Graduate University for Advanced Studies \\
           1-1, Ooho, Tsukuba, Ibaraki, 305-0801, Japan},
   email={shunzo.kumano@kek.jp}}

\author{N. Saito}{
  address={Department of Physics, Kyoto University,
           Kyoto, 606-8502, Japan},
   email={saito@nh.scphys.kyoto-u.ac.jp}}
   
\begin{abstract}
We report global analysis results for polarized parton distribution
functions in the nucleon. The optimum distributions are determined by
using spin asymmetry data on polarized lepton scattering on proton,
neutron, and deuteron. Their uncertainties are estimated by the Hessian
method. As a result, polarized quark distributions are relatively well
determined, whereas the polarized gluon distribution has a large uncertainty
band. We find that the obtained gluon distribution is compatible with recent
$\Delta g/g$ measurements in high-$p_T$ hadron productions.
\end{abstract}

\maketitle

\section{Introduction}

The nucleon spin is one of the fundamental quantities in physics, yet
we do not understand its origin. Internal structure of the nucleon
spin has been investigated mainly by polarized lepton-nucleon scattering
experiments. Polarizations of quarks and gluons are expressed by polarized
parton distribution functions (PDFs), and we extract them from the experimental
measurements.

The standard method is to obtain the polarized PDFs, which are 
expressed by a number of free parameters, by a $\chi^2$ analysis of available
polarization-asymmetry data. Among such analyses, we report the results
by the Asymmetry Analysis Collaboration (AAC) \cite{aac00, aac04}.
In the recent version \cite{aac04}, the polarized PDFs and their
uncertainties are obtained by using inclusive deep inelastic scattering (DIS)
data. However, these data are not enough to impose considerable constraints
on each polarized PDF, especially the gluon distribution.
Fortunately, the situation is changing because the RHIC-Spin started producing
data and there are reports on the gluon polarization by using semi-inclusive
lepton scattering data. In the following, we explain the AAC results
\cite{aac04} and also a comparison with recent $\Delta g/g$ measurements.

\section{Global analysis of spin asymmetry data}

For probing the internal structure of the nucleon, cross sections
for polarized deep inelastic lepton-nucleon scattering have been
measured. Their experimental data are shown by the spin asymmetry $A_1$.
It is expressed by the ratio of the polarized structure function $g_1$
to the unpolarized one $F_1$, which is usually expressed by
$F_2$ and the longitudinal-transverse ratio $R$:
\begin{equation}
   A_1(x, Q^2)=\frac{g_1(x, Q^2)}{F_2(x, Q^2)}\,
               2 \, x \, [1+R(x, Q^2)] \, .
\label{eqn:a1}
\end{equation}
The unpolarized PDFs of the GRV98 are used for calculating $F_2$,
and the SLAC parametrization is used for $R$. 
The polarized structure function is expressed in terms of polarized PDFs:
\begin{equation}
g_1 (x, Q^2) = \frac{1}{2}\sum\limits_{i=1}^{n_f} e_{i}^2
   \bigg\{ \Delta C_q(x,\alpha_s) \otimes [ \Delta q_{i} (x,Q^2)
    + \Delta \bar{q}_{i} (x,Q^2) ]
    + \Delta C_g(x,\alpha_s) \otimes \Delta g (x,Q^2) \bigg\},
\label{eqn:g1}
\end{equation}
where $\Delta C_q$ and $\Delta C_g$ are coefficient functions. 

A polarized PDF, $\Delta f$ ($=\Delta q_{i}$, $\Delta \bar{q}_{i}$,
or $\Delta g$), is defined by a number of parameters at the initial scale
$Q^2$=1 GeV$^2$: $\Delta f(x) = [\delta x^{\nu}-\kappa (x^{\nu}-x^{\mu})] f(x)$,
where $f(x)$ is the corresponding unpolarized distribution. This functional
form is convenient especially for imposing the positivity condition,
$|\Delta f(x)| \le f(x)$, in the global analysis. 
Flavor symmetric antiquark distributions are assumed at the initial scale
because of the lack of experimental information. The distributions are evolved
to experimental $Q^2$ points of the spin asymmetry by the DGLAP evolution
equations in order to calculate $\chi^2$ values at the same $Q^2$ points.
All the calculations are done in the next-to-leading order (NLO)
of $\alpha_s$ and the $\overline{\rm MS}$ scheme is used.
The parameters are determined so as to minimize
the total $\chi^2$. Then, the uncertainties of obtained PDFs are
estimated by the Hessian method. So far, only the inclusive DIS data
are used for the analysis in the AAC analysis.
Because there are measurements after the publication \cite{aac04},
an analysis is in progress with JLab, HERMES, and COMPASS data.
In future, we need to include other data such as the ones from
semi-inclusive DIS and RHIC-Spin. 

\section{Results}

\begin{wrapfigure}{r}{0.58\textwidth}
\begin{center}
\vspace{-0.4cm}\hspace{0.15cm}
\includegraphics[width=0.265\textwidth]{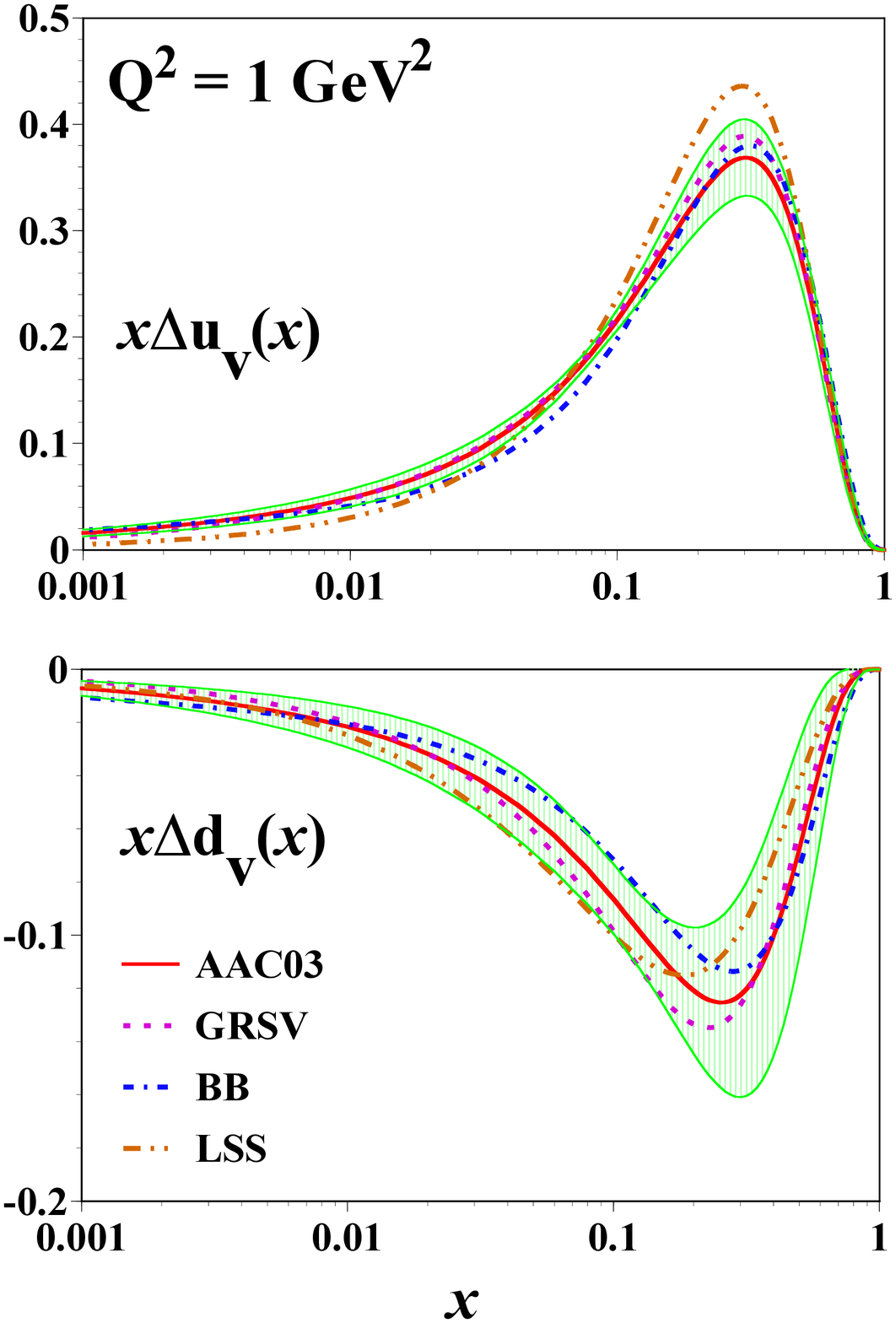}
\includegraphics[width=0.270\textwidth]{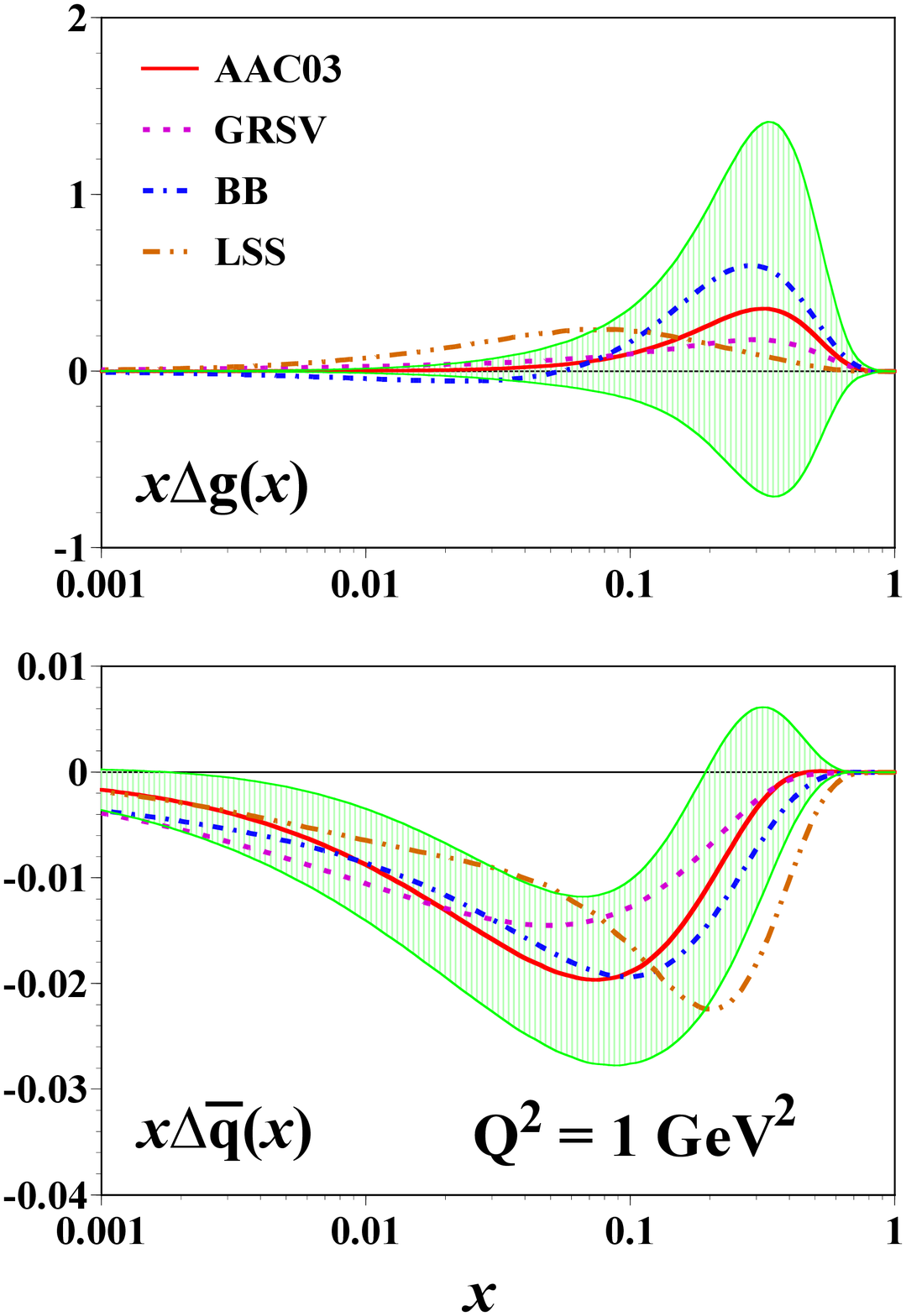}
\vspace{-0.3cm}
  \begin{minipage}{0.50\textwidth}
  \vspace{+0.2cm}
  {\footnotesize {\bf FIGURE 1.}
           Polarized PDFs with uncertainties \cite{aac04}.}
  \end{minipage}
\vspace{+0.2cm}
\end{center}
\end{wrapfigure}
We analyzed the data of the spin asymmetries $A_1$ in Eq. (\ref{eqn:a1})
for the proton, neutron, and deuteron in polarized lepton scattering.
The polarized PDFs obtained by the analysis at $Q^2$=1 GeV$^2$ are
shown in Fig. 1, where the solid curves indicate the AAC03
distributions and their uncertainties are shown by the bands.
In comparison, other analysis results (BB, GRSV, LSS)
are also shown in the figure. The valence-quark distributions are well
determined; however, the antiquark and gluon distributions have large
uncertainties. In particular, the polarized gluon distribution has
the huge uncertainty which is much larger than the distribution itself.
If the uncertainty is taken into account, even $\Delta g(x)=0$
is allowed at this stage. Although there are variations among
the four parametrizations in Fig. 1, they are consistent with
each other within the PDF uncertainties. From the PDF determination,
we find that the quark spin content is $\Delta \Sigma=0.21 \pm 0.14$
and the first moment of $\Delta g(x)$ is $\Delta g=0.50 \pm 1.27$
at $Q^2$=1 GeV$^2$.

\begin{wrapfigure}{r}{0.45\textwidth}
\begin{center}
\vspace{-0.3cm}\hspace{0.15cm}
\includegraphics[width=0.42\textwidth]{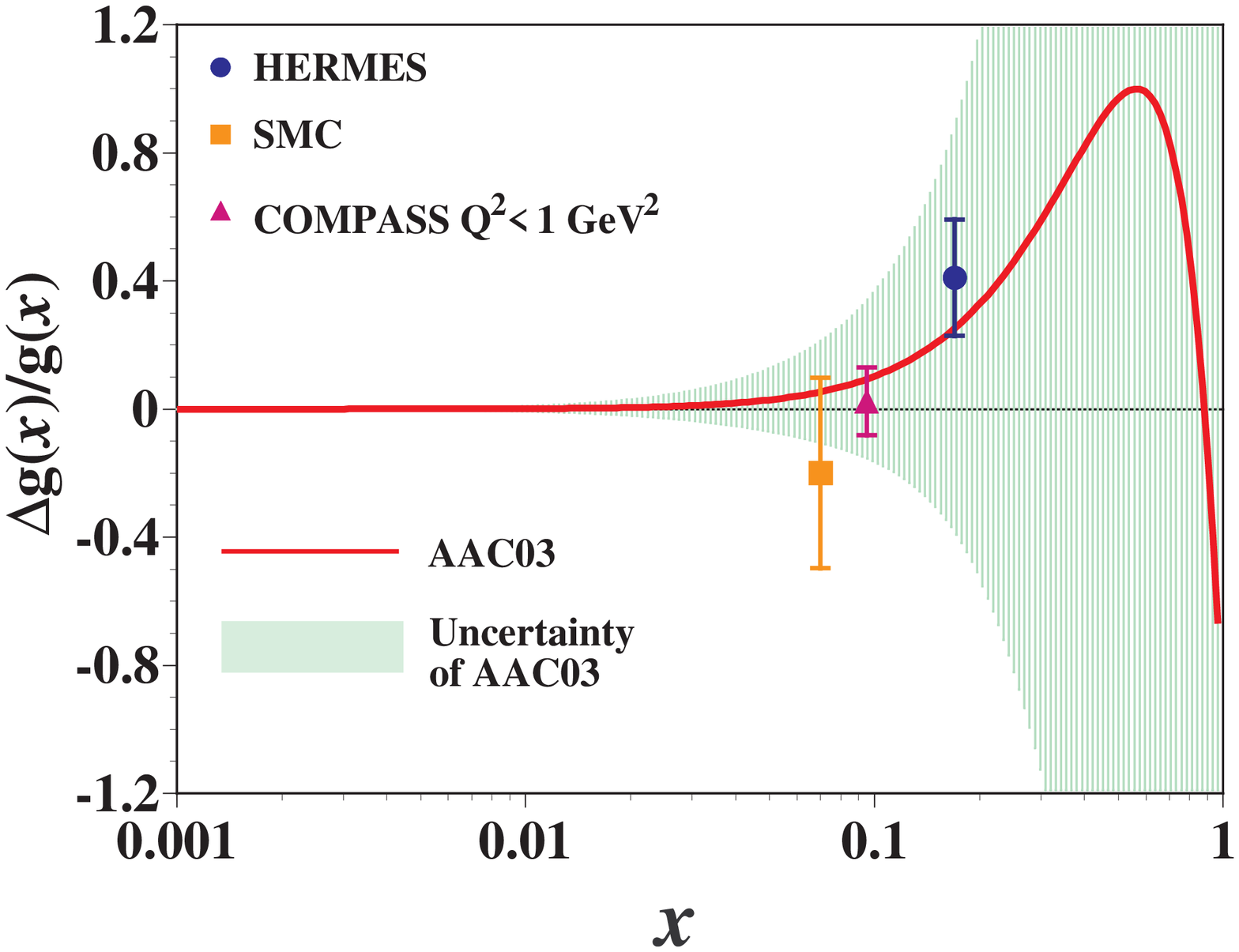}
\vspace{-0.3cm}
  \begin{minipage}{0.37\textwidth}
  \vspace{+0.2cm}
  {\footnotesize {\bf FIGURE 2.}
           AAC $\Delta g/g$ in comparison with high-$p_T$
           hadron-production data.}
  \end{minipage}
\vspace{0.1cm}
\end{center}
\end{wrapfigure}
The results indicate that it is impossible to determine a reliable
gluon polarization at this stage. There are two types of contributions
to $g_1$ from the polarized gluon distribution. One is through
the NLO coefficient function in Eq. (\ref{eqn:g1}), and the other
is through the DGLAP $Q^2$ evolution. Current inclusive DIS
experimental data are not accurate enough to specify these
gluonic effects, which results in the huge gluonic 
uncertainty. On the other hand, there are reports on the gluon
polarization by observing high-$p_T$ hadrons in DIS.
The HERMES, SMC, and COMPASS results are shown in Fig. 2 in comparison
with the AAC03 gluon distribution with the uncertainty
at $Q^2$=1 GeV$^2$.  There are also preliminary results
on $\Delta g/g$ from the COMPASS collaboration by using
large-$Q^2$ and charmed-meson data. We find that the AAC03 curve
agrees with three data points in Fig. 2.

After the AAC03 analysis, measurements were reported by JLab, HERMES,
and COMPASS collaborations on the spin asymmetry $A_1$.
In addition to these DIS measurements, high precision data of
$A_{LL}(\pi^0)$ from PHENIX and $A_{LL}({\rm jet})$ from STAR experiment
measured in polarized $pp$ collisions at RHIC are reported
at this conference. The AAC03 distributions are compatible with
the reported $A_{LL}(\pi^0)$. These new data should improve the situation
on the determination of polarized PDFs. In particular, the RHIC data
as well as the high-$p_T$ DIS data can impose considerable constraints
on the gluon polarization. 

The AAC polarized PDF code is available at the web site \cite{aacweb},
and it can be used for calculating the PDFs in a given kinematical
point of $x$ and $Q^2$. It is noteworthy that a code is provided
also for calculating the PDF uncertainties.

\begin{theacknowledgments}
\vspace{-0.20cm}
S.K. was supported by the Grant-in-Aid for Scientific Research from
the Japanese Ministry of Education, Culture, Sports, Science, and Technology.
This work is supported by the Japan-U.S. Cooperative Science Program. 
\end{theacknowledgments}



\end{document}